\documentstyle[12pt,preprint,aps]{revtex}

\tightenlines
\def\bold#1{\setbox0=\hbox{$#1$}%
     \kern-.025em\copy0\kern-\wd0
     \kern.05em\copy0\kern-\wd0
     \kern-.025em\raise.0433em\box0 }
\def\slash#1{\setbox0=\hbox{$#1$}#1\hskip-\wd0\dimen0=5pt\advance
       \dimen0 by-\ht0\advance\dimen0 by\dp0\lower0.5\dimen0\hbox
         to\wd0{\hss\sl/\/\hss}}
\newcommand{\be}{\begin{equation}}
\newcommand{\ee}{\end{equation}}
\newcommand{\bea}{\begin{eqnarray}}
\newcommand{\eea}{\end{eqnarray}}
\newcommand{\ba}{\begin{array}}
\newcommand{\ea}{\end{array}}
\newcommand{\nn}{\nonumber}

\begin{document}
\draft
\preprint{Napoli DSF-T-56/97}

\title{$^{}$ \\
$^{}$ \\
Stress-Tensor for parafermions from \\ 
the generalized Frenkel-Kac 
construction of affine algebras }

\author{Vincenzo Marotta$^{\dag}$}

\address{$^{\dag}$ 
Dipartimento di Scienze Fisiche,   
 Universit\'a di Napoli ``Federico II'' \\   
and INFN, Sezione di Napoli \\ 
Mostra d'Oltremare Pad.19-I-80125 Napoli  }

\maketitle

$\vspace{1.5cm}$

\begin{abstract}
I discuss a realization of stress-tensor for parafermion theories following  
the generalized Frenkel-Kac construction for higher level Kac-Moody 
algebras.
All the fields are obtained from $d$=rank free bosons compactified on torus.
This gives an alternative realization of Virasoro algebra in terms of a 
non-local correction of a free field construction which does not fit the usual 
background charge of Feigin-Fuchs approach. 
\end{abstract}

\vfill


\pagebreak

Two-dimensional conformal field theory is a subject of intensive study with 
applications in two-dimensional critical systems and string theory.  
The infinite-dimensional symmetry implied by the conformal invariance in two 
dimensions is encoded in the holomorphic and antiholomorphic components of 
stress-tensor, whose modes form two copies of Virasoro algebra. 

A lot of properties and techniques are now well known for these theories as 
the Feigin-Fuchs (FF) (or Coulomb gas) background charge method that can be 
applied to a free bosonic theory to decrease the value of central charge.   
This `bosonization' is based on the fact that a suitable combination of free 
fields with a stress-tensor at most quadratic in the fields can describe a 
large class of theories (including the minimal series of the Virasoro algebra).

An example of application of this technique is the well known $SU(2)/U(1)$ 
coset to the level $k$ that was extensively used to realize explicitly $Z_{k}$ 
parafermions \cite{ZF}. 
Parafermion algebras are generated by a collection of currents $\psi_{l}(z)$ 
and their conjugate $\psi_{l}^{\dagger}(z)$ with $\psi_{0}(z)=1$. A priori 
there are no restriction on the range of the integer $l$ and, generically, the 
conformal dimension $\Delta_{l}$ is fractional but, in all the case in which 
they are related to an affine extension of Lie algebras with $k<\infty$ of 
the kind considered here, they have a discrete symmetry that reduce the range 
of $l$ to a finite set. 

In general, only one parafermion (for instance $\psi(z)=\psi_{1}(z)$) and its 
conjugate are necessary to generate the full set of fields.

Nevertheless, how will be clear in the following, exists an alternative way to do 
it (at least for special values of central charge) by using a projection of 
the bosons on a multi-sheeted two-dimensional surface. 
This method comes from the study of Kac-Moody (KM) affine algebras to higher 
level in the conformal gauge. 
Using this result I show that a parafermion theory can be constructed 
factorizing the $d$=rank free bosons of the Cartan sub-algebra. 
While in the FF method a background charge to the infinity is used to deplete 
the conformal anomaly, in the present case the same effect is obtained by an 
additional term quadratic in the parafermions but evaluated in two different 
points.      

In this letter I will discuss the construction of stress-tensor for these 
systems where the parafermions are realized by means of factorization 
introduced in \cite{VM}.
I want to emphasize the difference between this approach and the approach of 
a recent paper \cite{GKN} where similar ideas are developed.
In \cite{GKN} no use is made of parafermions and the calculations are done 
for full affine algebra g using the mode algebra technique.
In the present paper, instead, I use the projection introduced in \cite{VM} 
to obtain the parafermions and the OPEs of these fields on the $k$-sheeted 
plane is directly obtained using a program of symbolic manipulation 
${\cal Mathematica}^{TM}$ \cite{TH,MAT}.
If the factorization of $U(1)^d$ terms is made in the expressions of 
ref.\cite{GKN}, the results become equals.
However, this factorization, except for the central charge, is not obvious.

The Goddard-Olive construction of affine KM algebras \cite{GO2}can be 
interpreted as a vertex construction of a Lie algebra in a singular lattice 
obtained by adding a light-like direction to the Euclidean lattice defining 
the horizontal finite Lie algebra.
The outer derivation is consistent with the extension of the singular lattice 
to a Lorentzian one and with the interpretation of the affine algebra as a 
sub-algebra of a Lorentzian one.
Let me shortly recall the essential steps of the covariant construction. 

I introduce the usual Fubini-Veneziano fields:
\be
Q^{\mu}(z)=q^{\mu}-ip^{\mu}lnz+i\sum_{n\neq 0} \frac{a^{\mu}_{n}}{n} z^{-n}
\label {eq: 4} 
\ee
and their derivatives $Q^{\mu(1)}(z)$ in a Minkowski space (I only consider 
holomorphic fields here).

If one considers the root $r$ belonging  to a Lorentzian lattice, it can be 
decomposed as $r=\alpha+nK^{+}+mK^{-}$ with $(K^{\pm})^2=0$, $K^{+}\cdot K^{-}=1$ 
and $\alpha$ belonging to $\Lambda$, the horizontal Euclidean lattice of any 
simply-laced Lie algebra. The affine sub-algebra is spanned by zero modes 
of the relevant generalized vertex operators:
\bea
A^{\alpha+nK^{+}}&=&\frac{c_{\alpha}}{2\pi i} \oint dz 
:e^{i \alpha+nK^{+}\cdot Q(z)}: \label{eq: MA} \\ 
H^{a}_{nK^{+}}&=&\frac{1}{2\pi i}\oint dz :Q^{a(1)}(z)e^{i nK^{+}\cdot Q(z)}
\label{eq: MH}\eea
$c_{\alpha}$ being a cocycle, and the $a$ index is restricted to the Euclidean 
lattice, with commutation relations:
\bea
\left[ H^{a}_{nK^{+}},H^{b}_{mK^{+}}\right]&=& n \delta^{ab} \delta_{n+m,0} 
K^{+}\cdot p  \\
\left[A^{\alpha+nK^{+}},A^{\beta+mK^{+}}\right] &=& 0 ~~~~~~~ \alpha \cdot 
\beta \geq 0  \\
\left[ A^{\alpha+nK^{+}},A^{\beta+mK^{+}}\right] &=& \epsilon (\alpha,\beta ) 
A^{\alpha+\beta+(n+m)K^{+}} \alpha\cdot \beta =-1  \\
\left[ A^{\alpha+nK^{+}},A^{-\alpha+mK^{+}}\right] &=& \alpha \cdot H_{(n+m)K^{+}} 
+n\delta_{n+m,0} K^{+}\cdot p   \hspace{1cm} \alpha=-\beta  \label {eq: 14}  \\
\left[ H^{a}_{nK{+}},A^{\alpha+mK^{+}}\right] &=& \alpha^{a} A^{\alpha+(n+m)
K^{+}} \label{eq: KM}
\eea

As $K^{+}\cdot a_{n}$ commutes with any element of this algebra,
it is possible to take them to be a constant and particularly:
\be
 K^{+}\cdot p \rightarrow k \hspace{1 cm} K^{+}\cdot a_{n} \rightarrow 0 
\hspace{1cm} if \:\:\:\: n \neq 0 \label {eq: 19}
\ee

In string theory this choice corresponds to a transformation by the covariant 
gauge to the transverse one. 
In this transformation the modes of eqs.(\ref{eq: MA},\ref{eq: MH}) become:
\bea
 A^{\alpha}_{n}&=&\frac{c_{\alpha}}{2 \pi i} 
\oint dz z^{nk} :e^{i \alpha\cdot Q(z)}: \\  
 H^{a}_{n}&=&\frac{1}{2\pi i} \oint dz z^{nk} Q^{a(1)}(z) 
\eea
and realize the same KM algebra but to the level $k$.

Energy-momentum stress-tensor for this algebra is obtained by means of the 
famous Sugawara construction \cite{SG} and supplies a Virosoro algebra 
realization with central charge $c=\frac{k dim 
g}{k+h^{\vee}}=\frac{dk(1+h^{\vee})}{k+h^{\vee}}$, where $h^{\vee}$ is the 
dual Coxeter number of Lie algebra $g$.

For $k>1$ the charge is greater then rank, so it seem to be impossible 
to realize the algebra with only $d$ Fubini fields as can be done for $k=1$ by 
means of the standard Frenkel-Kac-Segal construction \cite{FKS}.
Parafermions was introduced by Gepner \cite{GQ} as auxiliary fields to take in 
to account this fact.

These theories contain fields of fractional spin and which are non-local.  
A special case are self-dual $Z_{k}$ models of Zamolodchikov and Fateev 
\cite{ZF,GQ} that are related to $SU(2)$ current algebra.

Generalized parafermions theories are labeled by the level $k$  and by a simple 
Lie algebra and have a discrete symmetry group which in the simply-laced case 
is $\Lambda/k\Lambda$ and central charge 
$c_{\psi}=\frac{dh^{\vee}(k-1)}{k+h^{\vee}}$.

Now it is possible to discuss the connection of this construction with the 
parafermion one.

For $k>1$ the Fock space $F^{k}$ of Heisenberg sub-algebra is build up by the 
set of $ H^{a}_{n} $ operators, which is the subset of the creation 
and annihilation operators of the whole Fock space $F $, $a^{\mu}_{m}$, with 
$\mu = a $ and $ m = nk $ with $ k $ fixed, while parafermions are constructed 
on the space $\Omega^{k}$, the vector space of vacuum vectors for the 
Heisenberg sub-algebra. 

To realize the parafermion coset one needs to factorize the $d$ Fubini 
fields $X^a(z)$ corresponding to the Cartan sub-algebra,   
this can be done defining the fields $\phi^{a,j}(z)=
Q^a(\varepsilon^{j}z)-X^a(z)$, where $\varepsilon = e^{\frac{2\pi i}{k}}$.
These fields transform as $d(k-1)$ bosons with twisted boundary conditions 
$\phi^{a,j+1}(z)=\phi^{a,j}(\varepsilon z)$. 

Moreover, the parafermions are defined on $\Omega^{k}$ by
\be
\psi^{\alpha}(z)=\frac{z^{\frac{1}{k}-1}}{k}\sum_{j=1}^{k}\varepsilon^{j}
:e^{i\alpha\cdot \phi(\varepsilon^{j}z)}: \label {eq: 30} 
\ee 
where $\alpha\cdot \phi(\varepsilon^{j}z)=\alpha\cdot \phi^{j}(z)$ and satisfy 
the fundamental product:
\be
\psi ^{\alpha}(z) \psi ^{\beta}(\xi )=
\frac{z^{\frac{1}{k}-1}\xi^{\frac{1}{k}-1}}{k^2} 
\sum_{j,j'=1}^{k}\varepsilon^{j+j'}
:e^{i\alpha\cdot \phi(\varepsilon^{j'}z)}
e^{i\beta\cdot \phi(\varepsilon^{j}\xi)}:
\frac{(\varepsilon^{j'}z-\varepsilon^{j}\xi)^{\alpha
\cdot \beta}}{(z^{k}-\xi^{k})^ {\frac{\alpha \cdot \beta}{k}}} 
\label {eq: 31}  
\ee

These fields live on the $k$-sheeted complex plane which is the image space of 
the conformal transformation $z^{k}\rightarrow z$, thus, each value of the 
variable $z^{k}$ corresponds to $k$ different points on the plane related by 
a discrete transformation. They can be interpreted as the parafermion fields 
for the $k$ level KM algebra.

I define also an isomorphism between single-value fields on the $k$-sheeted 
complex plane and multi-values fields on the one-sheeted plane 
by means of the following identifications:
\be
a^{a}_{nk+l} \longrightarrow \sqrt{k}a^{a}_{n+l/k} \hspace{1cm} q^{a} 
\longrightarrow \frac{1}{\sqrt{k}}q^{a} \label {eq: 32}
\ee

All the physical fields obtained by this isomorphism are single-valued by 
construction.
In the following the fields on one-sheeted plane are denoted with an hat on 
the same symbol corresponding to the fields on multi-sheeted complex plane.

By the definition eq.(\ref {eq: 30}) and by eq.(\ref {eq: 31}) 
using eqs.(\ref {eq: 32}) and the conformal transformation 
$z^{k}\rightarrow z$ one obtain the parafermion OPE relations for 
$\alpha \cdot \beta=-1 $:
\be
\hat{\psi}^{\alpha}(z) \hat{\psi}^{\beta}(\xi)= (z-\xi)^{-1+\frac{1}{k}} 
\left [\hat{\psi}^{\alpha+\beta}(\xi) + {\cal O}(z-\xi)\right] \label {eq: 39}
\ee
and for  $ \alpha =- \beta $
\be
\hat{\psi}^{\alpha}(z) \hat{\psi}^{- \alpha}(\xi)= (z-\xi)^{-2+\frac{2}{k}}  
\left [1 + {\cal O}(z-\xi)^2\right ] \label {eq: 38} 
\ee

As it is well known a duality exists between bosonic theories with 
compactification radius R and 1/R and the resulting theories are recognized to 
be completely equivalent.
The transformation here defined acts just as a duality that reduce the 
compactification radius $\sqrt{k}$ to $1/\sqrt{k}$ but, while for the usual 
single-value case no further affects arise, in the multi-value case a mapping 
between multi-local and non-local terms appears.
This can be very useful to perform computations with terms that should be very 
complicated on the one-sheeted plane.
  
The boundary conditions are selected by the discrete symmetry derived from the 
charges in the coset $\Lambda ^{*}/k\Lambda $ which is imposed in order to 
have a single-valued integrand.

Let me emphasize once more that the parafermions appear naturally in this 
procedure and they are built on the bosonic space by means of fields through 
eq.(\ref {eq: 30}), so one can consider the present procedure to obtain 
parafermions as a generalized bosonization procedure.

Now it is possible to use the exact formula eq:(\ref{eq: 31}) to define the 
stress-tensor for generalized parafermions.
 
Single value stress-tensor field $T_{\psi}(z)$ is defined as the normally 
ordered product of two parafermions on the $k$-sheeted complex plane:
\be
T_{\psi}(z)= \frac{c_{\psi}}{d h^{\vee} \Delta_{\psi}} 
\oint \frac{dw}{2 \pi i} \frac{w^{k-1}}{2 k^2}\sum_{\alpha\in\Lambda}
\sum_{j,j'=1}^{k}\frac{\varepsilon^{j+j'}
:e^{-i\alpha\cdot \phi(\varepsilon^{j'}w)}
e^{i\alpha\cdot \phi(\varepsilon^{j}\xi)}:}{(w^k-z^k)}  
\ee
where $c_\psi$ is the central charge and $\Delta_\psi$ the conformal weight of 
parafermions \cite{G}.

Using the product eq.(\ref{eq: 31}) and writing the function 
$(w^k-z^k)$ as $\prod_{l=1}^{k}(w-\varepsilon^{l}z)$ the above expression 
can be written as the sum of
\be
T_{L}(z)=\frac{h^{\vee}}{(k+h^{\vee})}\left(
-\frac{1}{2 k^2}\sum_{j=1}^{k}
:\partial_{z}\phi^{j}(z)\cdot  \partial_{z}\phi^{j}(z): + 
\frac{d(k+1)}{24 k z^2}\right)
\ee
coming from the $k-1$ poles of order two in 
$z_1=\varepsilon^{j}z$ and containing the usual term in the derivatives of the 
fields $\phi$ that give rise to a central charge $c=d(k-1)$ (but with a diverse 
normalization), and 
\be
T_{Q}(z)=\frac{c_{\psi}}{d h^{\vee} \Delta_{\psi}}  
\frac{1}{2 k^2}
\sum_{\alpha\in\Lambda}\sum_{j'\neq j=1}^{k}
\frac{\varepsilon^{j+j'}
:e^{-i\alpha\cdot \phi(\varepsilon^{j'}z)}
e^{i\alpha\cdot \phi(\varepsilon^{j}z)}:}{k z^2  
(\varepsilon^{j'}-\varepsilon^{j})^2}  
\ee

The new features of these relations are related to the $T_{Q}(z)$ term which 
is due to the additional poles appearing in the not-coincident points and that 
modify strongly the OPE.  
This is a controllable form of non-locality of parafermion product that 
allows to compute easily OPE that should be very complicated on the 
one-sheeted plane.

OPE of $T(z)$ fields follows by the three exact product formulas:
\bea
T_{L}(z)T_{L}(\xi)=&N_k&
\sum_{\alpha ,\beta\in\Lambda}
\sum_{j_{1},j_{2}=1}^{k}\frac{\varepsilon^{2(j_{1}+j_{2})}}{4 k^4}
\frac{\partial^{2}_{z_{1}\rightarrow \varepsilon^{j_{1}}z}}{2}
\frac{\partial^{2}_{\xi_{1}\rightarrow \varepsilon^{j_{2}}\xi}}{2}  
\frac{z_{1}^{k-1}\xi_{1}^{k-1} 
{\cal O}^{(-\alpha,\alpha,-\beta,\beta)}
(z_1,\varepsilon^{j_1}z,\xi_1,\varepsilon^{j_2}\xi)}{
\prod_{i=1}^{k-1}(z_{1}-\varepsilon^{i+j_1}z)
\prod_{i=1}^{k-1}(\xi_{1}-\varepsilon^{i+j_2}\xi)}  \nn \\ 
&\times&\left(\frac{z_{1}-\xi_{1}}{z_{1}-\varepsilon^{j_{2}}\xi}
\frac{\varepsilon^{j_{1}}z-\varepsilon^{j_{2}}\xi}
{\varepsilon^{j_{1}}z-\xi_{1}}\right)^{
\alpha\cdot\beta}
\left(\frac{z_{1}^k-\xi_{1}^k}{z_{1}^k-\xi^k}
\frac{z^k-\xi^k}{z^k-\xi_{1}^k}\right)^{
-\frac{\alpha\cdot\beta}{k}}
\eea

\bea
T_{L}(z)T_{Q}(\xi)=&N_k&
\sum_{\alpha ,\beta\in\Lambda}
\sum_{j_{1},j'_{2}\neq j_{2}}^{k}\frac{\varepsilon^{2 j_{1}+j_{2}+j'_{2}}
}{4 k^4} 
\frac{\partial^{2}_{z_{1}\rightarrow \varepsilon^{j_{1}}z} }{2}
\frac{z_{1}^{k-1} 
{\cal O}^{(-\alpha,\alpha,-\beta,\beta)}
(z_1,\varepsilon^{j_1}z,\varepsilon^{j'_2}\xi,\varepsilon^{j_2}\xi)}{k \xi^2 
\prod_{i=1}^{k-1}(z_{1}-\varepsilon^{i+j_1}z)
(\varepsilon^{j'_2}-\varepsilon^{j_2})^2} \nn \\ 
&\times&\left(\frac{z_{1}-\varepsilon^{j'_2}\xi}{z_{1}-\varepsilon^{j_{2}}\xi}
\frac{\varepsilon^{j_{1}}z-\varepsilon^{j_{2}}\xi}
{\varepsilon^{j_{1}}z-\varepsilon^{j'_2}\xi}\right)^{\alpha\cdot\beta}
\eea

\bea
T_{Q}(z)T_{Q}(\xi)=&N_k&
\sum_{\alpha ,\beta\in\Lambda}
\sum_{j'_{1}\neq j_{1},j'_{2}\neq j'_{2}}^{k}
\frac{\varepsilon^{j_{1}+j_{2}+j'_{1}+j'_{2}}}{4 k^4} 
\frac{{\cal O}^{(-\alpha,\alpha,-\beta,\beta)}
(\varepsilon^{j'_1}z_1,\varepsilon^{j_1}z,
\varepsilon^{j'_2}\xi,\varepsilon^{j_2}\xi)}{k^2 z^2 \xi^2 
(\varepsilon^{j'_1}-\varepsilon^{j_1})^2
(\varepsilon^{j'_2}-\varepsilon^{j_2})^2} \nn  
\\ 
&\times&\left(\frac{\varepsilon^{j'_1}z-\varepsilon^{j'_2}\xi}
{\varepsilon^{j'_1}z-\varepsilon^{j_{2}}\xi}
\frac{\varepsilon^{j_{1}}z-\varepsilon^{j_{2}}\xi}
{\varepsilon^{j_{1}}z-\varepsilon^{j'_2}\xi}\right)^{
\alpha\cdot\beta}
\eea 
where I define the four points vertex operator:
\be
{\cal O}^{(\alpha,\beta,\gamma,\delta)}(z_1,z,\xi_1,\xi)=
:e^{-i\alpha\cdot \phi(z_1)}
e^{i\beta\cdot \phi(z)}
e^{-i\gamma\cdot \phi(\xi_1)}
e^{i\delta\cdot \phi(\xi)}:
\ee
and $N_k=\left(\frac{c_{\psi}}{d h^{\vee} \Delta_{\psi}}\right)^2=
\frac{k^2}{(k+h^{\vee})^2}$.

All the above terms give rise to analytical functions with poles in 
$z=\varepsilon^{l}\xi$ for $l=1,...,k$. 
To evaluate these OPEs one needs to expand the products on the multi-sheeted 
plane, then applying the conformal transformation $z^k\rightarrow z$, one 
obtain the OPEs on single-sheeted complex plane. 
While $T_{Q}(z)$ term is multi-local on this plane the transformed operator 
becomes non-local on one-sheeted plane.

Virasoro algebra is defined as the modes algebra of $\hat{T}(z)$ stress-tensor 
and can be obtained by means of the usual Laurent expansion in the $z$ 
variable.
Central charge computed by means of this technique takes a contribution by  
$\hat{T}_{L}(z)$, $c_{L}= \frac{d (h^{\vee})^2 (k-1)}{(k+h^{\vee})^2}$ 
which can be recognized as the usual charge of $d(k-1)$ bosons with a 
diverse normalization, but to give the correct value of charge of the 
generalized parafermions system an additional term due to $\hat{T}_{Q}(z)$ is 
generated $c_{Q}= \frac{d h^{\vee} k (k-1)}{(k+h^{\vee})^2}$.  

This completely new form of stress-tensor is unrelated to FF method where a 
deformation in the second derivatives of $\phi$ is added to the standard 
$T_{L}(z)$ due to the coupling with the background charge.
Moreover, in the present case, deformation comes by the projection on an 
algebraic curve of the system of $\phi$ bosons and has topological origin.   

Notice also that the products in this formalism are exactly defined, thus 
the additional not polynomial term $T_{Q}(z)$ is of non-perturbative kind and 
cannot follow by any continuous deformation of the standard free fields 
realization as happen for FF method.

Also the action of stress-tensor on a parafermion can be evaluated following 
the present procedure by using of the master product of eq:(\ref{eq: 31}) and 
gives the exact value of conformal weight only by adding of $T_{Q}(z)$ 
contributions.

Let me emphasize once more that although the final expression it is similar 
to previous known ones \cite{BT,H} their origin are completely different.
For instance, in this approach the discrete symmetry of parafermions is a 
direct  consequence of the geometry of the k-sheeted plane, on the contrary, 
in the usual realizations this must be imposed to the $d(k-1)$ independent 
bosons for construction.
Instead, in the present case, all the relevant fields are, naturally, obtained 
by means of a suitable projection procedure applied to the unique really 
independent field $Q(z)$ while the bosons $\phi^j(z)$ are not truly 
independent but only evaluated in diverse positions. 

Following the arguments in the last part of \cite{G} one can apply these 
results also to realize not-simply laced algebras to the level $k=1$. 
For instance, by the equivalence between the parafermions used in the case of 
simply-laced algebras and the auxiliary fields for level $k>1$ 
(for example see. \cite{BT}) that needs to associate to the short roots one 
find the equivalence: 
$C_{n},k=1\approx A_{n-1},k=2$; $B_{n},k=1\approx A_{1},k=2$; 
$F_{4},k=1\approx A_{2},k=2$; $G_{2},k=1\approx A_{1},k=3$; 

Parafermions arise naturally also in the context of the $N=2$ superconformal 
algebra.
Its generators, $J(z)$, $T(z)$ and $G^{\pm}(z)$ can be constructed in terms of 
one scalar field and two parafermion currents.
In the particular case of central charge less then $3$, the parafermions are 
the same fields appearing in the $SU(2)_k$ model.
This correspondence allow to obtain all the unitary representations of 
superconformal algebra from the representations of $SU(2)_k$ current algebra by 
subtracting and the adding back a free boson.
It is evident that this procedure can be performed also in the present case by 
using the parafermions introduced in this letter.

Of course one of the most interesting application of the covariant vertex 
construction is the realization of Lorentzian algebras \cite{GO,MS,B}, in fact, 
the  unified construction of arbitrary level representations of affine KM 
algebras appears quite naturally in this context, where the level $ k $ can 
be changed by the action of the pure Lorentzian generators, in complete 
analogy with the case of affine algebras where the weights of horizontal 
finite dimensional Lie sub-algebra are changed by the action of the affine 
generators.

It should also be interesting to study in detail the representation theory 
for these coset and the existence of a form of screening charge and null 
vectors, to understand more deeply the structure and the connections with 
Lorentzian algebras which, recently, have been recognized to be related to 
non-perturbative effects in string theory \cite{N,HM}.

\bigskip

{\bf Acknowledgments} - The author is indebted to A. Sciarrino for useful 
comments and the reading of the manuscript.

\end{document}